# Fourier Transform of Annular Beams

Dimo N. Astadjov

*Metal Vapor Lasers Department, Georgi Nadjakov Institute of Solid State Physics, Bulgarian Academy of Sciences, 72 Tzarigradsko Chaussee, Sofia 1784, Bulgaria*   asta55@issp.bas.bg

(Dated: 13 April 2009)

Fourier transform is applied to annular beams of simplified flat two-level geometry: bright outer ring with a darker core. The pattern of focal beam profile (i.e. far field) is calculated and characterized with respect of its intensity structure. Specific behavior is observed for extreme annularity values.

Many applications of lasers are closely related and determined by the properly chosen beam profile. Inherently laser-produced radiation has a great diversity of profiles. These raw profiles are not always good enough for straight use and they need to be tailored. Last decade features an enormous interest in the laser beam shaping. Abundance of techniques and models occur that can produce profiles required for the great variety of laser applications. In 2000 appeared a comprehensive reference book [1] which summarized the most fundamental theories and techniques and provided all the basic information to research, develop, and design beam shaping systems.

Copper vapor lasers are the most high-power and high-efficiency lasers which produce visible light in a straightforward way without any proceeding light conversion. Their beam profile shape varies from annular or top-hat to Gaussian-like. In the case of MOPA CuBr laser profile variations are observed in experiments on timing and buffer gas composition. While annular profiles are typical for non-hydrogen buffer, top-hat profiles are predominant for buffers containing hydrogen [2, 3]. For many uses the laser radiation is concentrated by focusing. The focusing causes a near-field beam profile to transforms into a far-field beam profile of intensity distribution often quite different from the initial near-field intensity distribution. Relations between near-field and far-field beam profiles (distributions) are of expanding interest due to the vast employment of optical (laser) beams in science and technology.

The two-dimensional far-field beam profiles can be calculated by the two-dimensional Fourier transform of the near-field beam profiles. In real physical world, this transform is easily performed by optical lenses. The focal spot profile is the far-field profile of near-field profile in front of the lens (within the accuracy of a phase coefficient) [4]. The Fourier transform is a very powerful tool for optical signal processing. The most popular and simple form is the Fast Fourier Transform (FFT). The FFT is applicable to the complex optical field. The signal that is optically detectable is the intensity of optical field. To find it, we have to calculate the product of the optical field complex amplitude and its conjugated optical field complex amplitude [5].

**Near- and far-field parameters of flat two-level annular beam**

The simplest annulus type is the normalized flat two-level annulus: the inside concentric circular area is of intensity, I1, which varies from 0 to 1 and the annulus itself is of intensity, I2=1. If I1=I2=1, the profile is top-hat; if I1=0, the profile is 'pure' annular. The pattern (of axial section) of annular near-field beam profiles can be described by two parameters (Fig. 1). The first parameter 'annularity', k is k=d1/d2, where d1 and d2 are the inside and the outside diameters of annulus. The second parameter, Idip is the intensity dip of the central area: Idip=I2-I1.

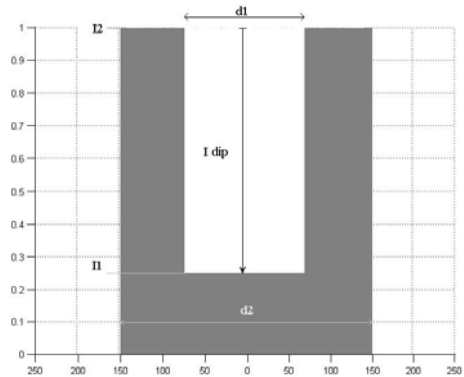

FIG.1: Intensity axial section of normalized two-level annulus; d1 and d2 are the inside and the outside diameters of annulus; the circular core area is of intensity, I1 (varying from 0 to 1), and I2=1.

The type of calculated pattern of far-field beam profiles is shown in Fig. 2. The profile can be described as having a prominent central peak and side peaks concentrically surrounding it (better seen in the figure inset). The appearance of side peaks comes from the discrete form of calculations done by FFT. Actually they should smear into uninterrupted concentric rings.

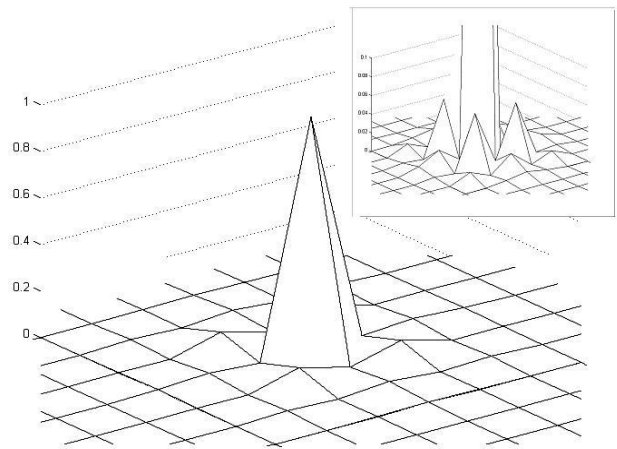

FIG.2: Typical calculated pattern of far-field beam profiles of a normalized two-level annulus; the inset is x10-magnified central area.

In Fig.3 we give the patterns of far field calculated from near fields of different values of k. (Z-axis is 10% of central peak intensity for better view of side peaks structure.) As k goes up so do the side peaks. For k=0 (top-hat) the near-field product of FFT is an Airy pattern of the far-field profile.

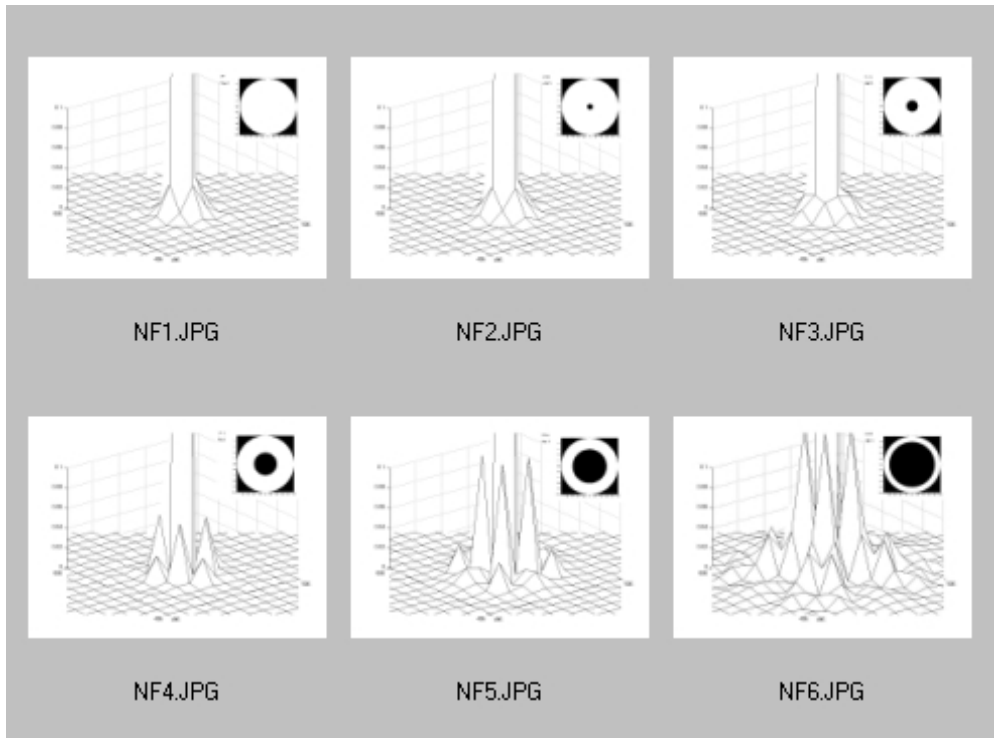

FIG.3: Patterns of far fields calculated from near fields of different values of annularity.

The far-field profile parameters specified here are with regard to laser beam utilization in applications which are not susceptible to phase components of light. We pay attention to the power (energy) component of light. Nevertheless the coherency of light is a requirement for all our simulations. As a major parameter we introduce the fraction of the central peak energy to the whole energy of beam, PF0. This parameter gives a notion about energy spread within the far-field spot. The higher PF0 the lower energy spread is. In practice that means less affected surrounding (the central peak) area by light radiation. The dependence of PF0 on Idip is plotted in Fig.4.

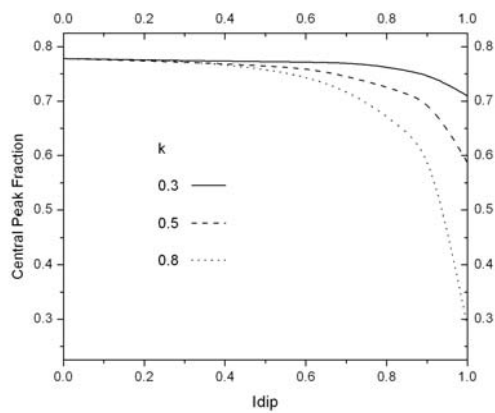

FIG.4: Dependence of the central peak energy fraction (PF0) on Idip (the intensity dip of the central area: Idip=I2-I1) for three values of annularity, k - 0.3, 0.5 and 0.8.

As can be seen the increase of Idip leads to the decrease the central peak energy fraction. At higher k the decrease starts at lower Idip. So at k=0.3 and Idip<0.75, the change of PF0 is less than

1.5%. While the maximum of PF0 is 0.778, for Idip=0.75 the fraction of the central peak energy PF0 is 0.767. But at k=0.8 for the same Idip of 0.75, PF0 is 0.7. The central peak energy fraction dominates the far-field energy distribution nearly over the whole range. The exception is for Idip>0.9 (and k>0.65) which occurs seldom in usual practice. Taking into account that the central peak energy is concentrated on a smaller spot area, the net impact of side energy spread diminishes furthermore.

For the sake of usability, a formula of PF0=f(Idip,k) of a fairly good approximation was calculated [6] and is given below:

$$z = a(\sinh(x)y^{1.5}) + b(\tan(x)y^{1.5}) + c$$

where a = 2.0310451454356748E+00, b = -1.9232938348115700E+00 and
c = 7.8402321357391169E-01; consider that $z \equiv PF0$, $x \equiv Idip$ and $y \equiv k$.

**Simulation of pure annular beams as dependents of the annularity parameter**

During the FFT simulation of annular beams of Idip=1 ('pure', 'dark', etc.) interesting correlations between certain parameters of the far-field intensity profile of these beams were found. Here we shall define some parameters which concern energy relations in the central area of the far-field intensity profile of these beams. So PF0 (the power fraction of "zero" center) is the magnitude of the highest intensity pixel divided by the sum of the values of all pixels of far-field intensity profile. We regard it as the power fraction of the very peak of the far-field intensity profile with respect to whole beam power. PF1 (the power fraction of "zero+1" center) is defined same way but here numerator yet includes the sum of the intensities of nearest 8 pixels around the peak intensity pixel (the "zero" center) plus the "zero" center itself (3x3 matrix). Also we can make use of the quantity pf=PF0/PF1, which actually is the magnitude of the highest intensity pixel over the sum of the intensities of nearest 8 pixels around it plus the peak pixel itself. All three quantities are plotted in Fig.5 as functions of beam annularity. As seen PF0 and PF1 go down with the increase of k; then they plunge for k>0.9 when k approaches unit (see the inset). Meanwhile the behavior of pf can be depicted as having two stable constant values which pf tends to. So in the left part of the line, pf=0.85 when k<0.25; and in the right end, pf=0.550 for k>0.94. These two points divide the pf-curve into three sections. The behavior of far-field 3D pattern is quite distinctive in the different sections. As a whole, the far-field 3D pattern changes in form and amplitude. This is shown in Fig.6, where for explicitness we fix the z-axis and feeble patterns (as of k≥0.8) are magnified.

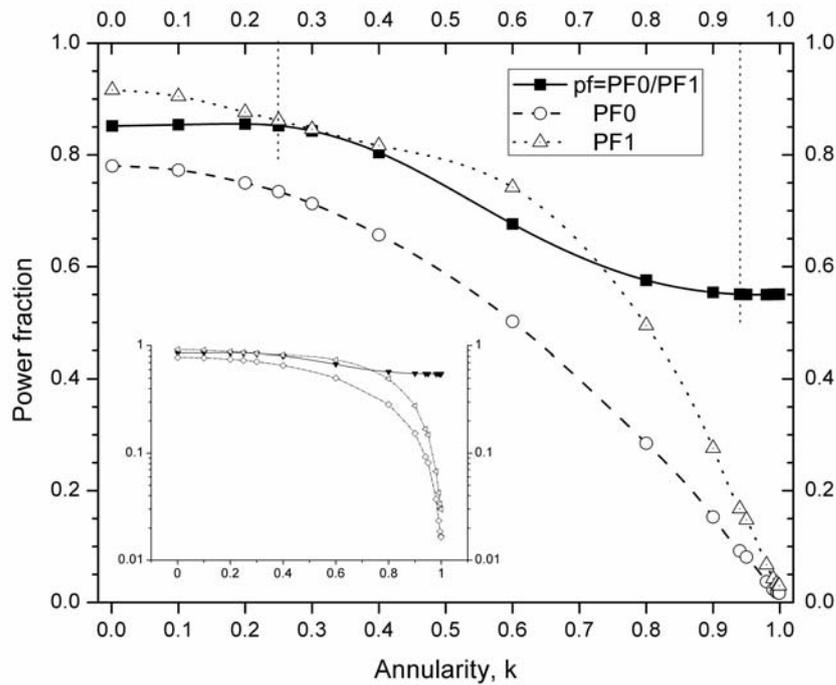

FIG.5: Power fraction parameters PF0, PF1 and pf as functions of beam annularity of pure (dark) annular beams.

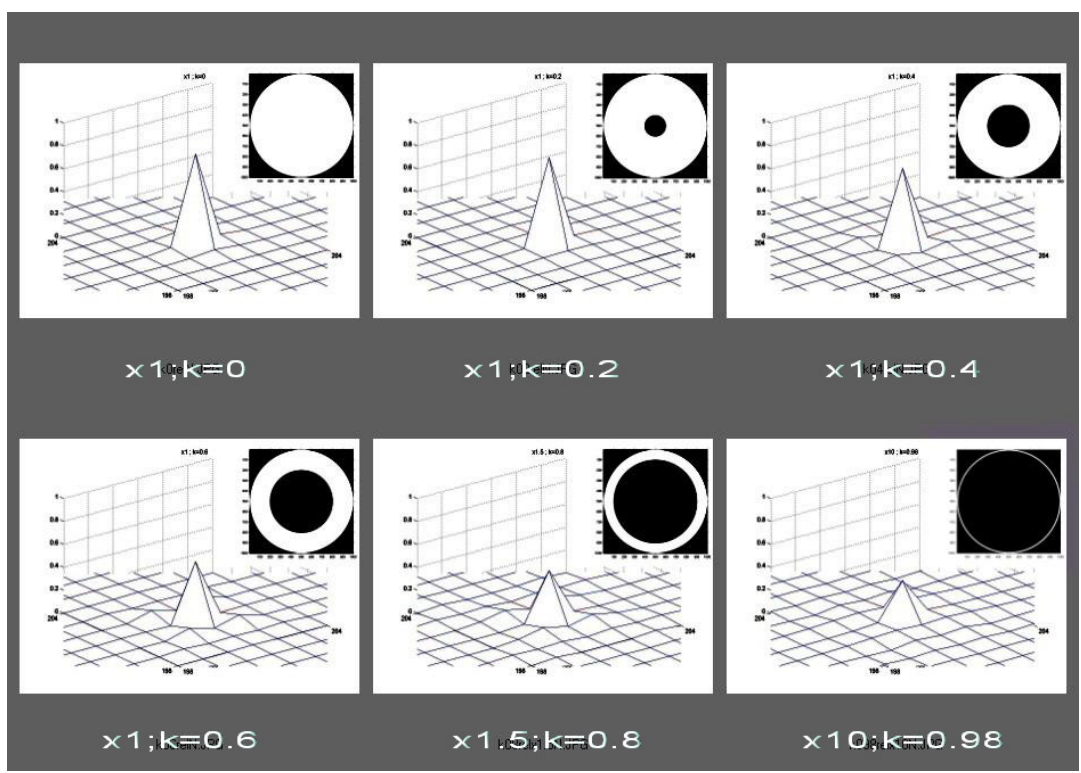

FIG.6: Far-field patterns of pure annular beams of different annularity; patterns of k≥0.8 are magnified.

Far-field 3D patterns in four points of the left section (k≤0.25) and four more of the right section (k≥0.94) of the pf-line are given in Fig.7 and Fig.8, respectively.

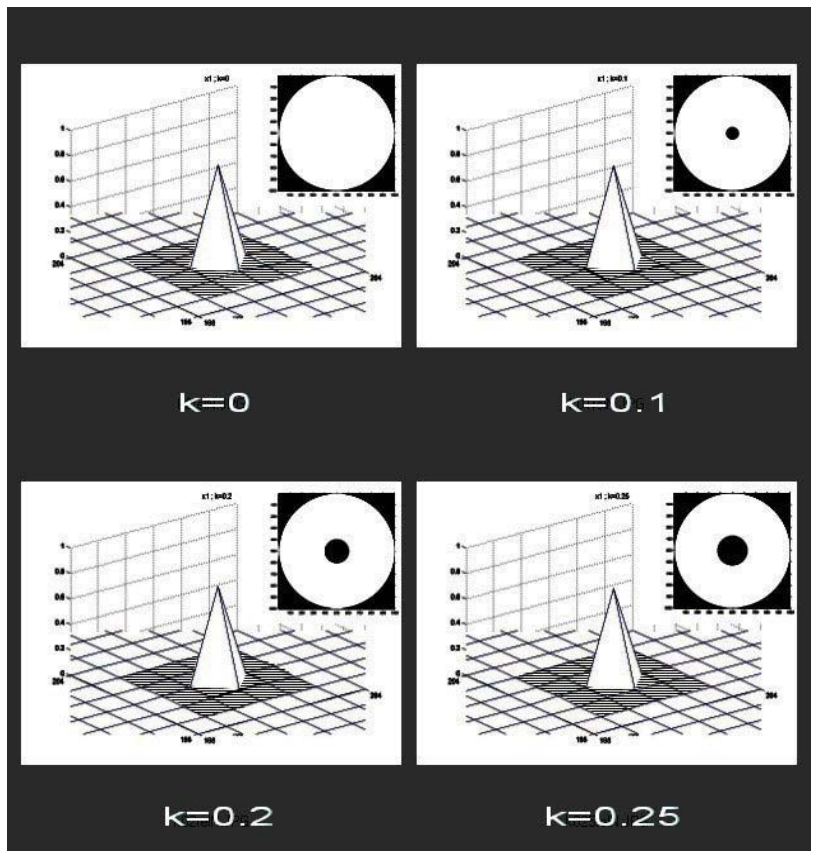

FIG.7: Far-field patterns of pure annular beams of annularity k≤0.25.

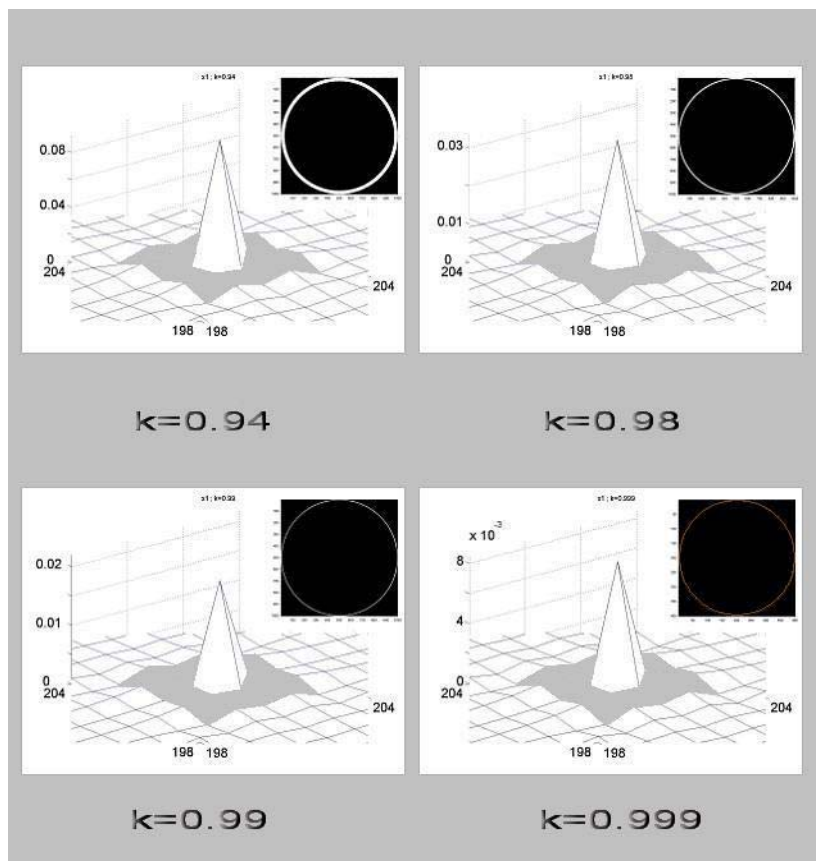

FIG.8: Far-field patterns of pure annular beams of annularity k≥0.94.

Despite the changes in near fields (the insets) we do not observe considerable variations of far-field structures. The only exception is the magnitude of the "zero" peak amplitudes of the right section of the pf-line (Fig.8). So, for k=0.999 the amplitude (i.e. PF0) is an order of magnitude less than for k=0.94 (which is PF0≈0.09).

**Discussion**

The simulation findings can be commented from several points of view. If we are interested of the technology-relevant power laser applications the simulation of beam annularity effect can give us a notion of the spread structure of focused light in the focal plane. As simulation showed serious malformation (say, the central peak fraction in Fig.4 becomes less than 0.5) of the far-field pattern is probable when the near-field annularity is high (k>0.65) and the annulus is almost pure (Idip>0.9 i.e. the central sag intensity is less than 10% of the ring intensity). This situation does not occur frequently. However, even in that case, since the central peak energy is concentrated into a smaller spot area, the net impact of side energy spread could not be so detrimental.

A quite unexpected result was the simulation finding that up to an annularity of 0.25 the far-field intensity pattern is actually not disturbed by the black hole in the near-field center (Fig.7). A parameter, pf was introduced as characteristic of different kinds of behavior of far-field intensity structure (Fig.5). The range of pf is 0.55≤pf≤0.85. So if pf=0.85 we have a steady-state far-field pattern produced from near field with a "small" (k<0.25) black hole in the center. The lower limit (0.55) of pf also corresponds to a steady-state kind of far-field pattern. This time near field is actually a thin annulus of k>0.94. The shape of the far-field structure is invariable but its amplitude diminishes with the increase of k. Having in mind that very thin annular beams are required for generation of Bessel beams [7], it is quite possible that k>0.94 is a reflection of a kind of threshold conditions for Bessel beam production.

This work was supported by Ministry of Education and Science of Bulgaria by Grant BIn3/07.